\documentclass{moriond}

\usepackage{graphicx}
\usepackage{color}

\def\be{\begin{equation}}
\def\ee{\end{equation}}
\def\bea{\begin{eqnarray}}
\def\eea{\end{eqnarray}}

\usepackage{xspace}
 \usepackage{ulem}

\newcommand{\Collier}{{\rmfamily\scshape Collier}\xspace}
\newcommand{\Munich}{{\rmfamily \scshape Munich}\xspace}
\newcommand{\Sherpa}{{\rmfamily\scshape Sherpa}\xspace}
\newcommand{\OpenLoops}{{\rmfamily\scshape OpenLoops}\xspace}

\newcommand{\OpenLoopsSherpaMunich}{{\rmfamily \scshape OpenLoops+Sherpa/Munich}\xspace}

\def\mathswitchr#1{\relax\ifmmode{\mathrm{#1}}\else$\mathrm{#1}$\fi}

\newcommand{\jet}{j}

\newcommand{\ord}{\mathcal{O}}

\newcommand{\HTtot}{H_{\mathrm{T}}^{\mathrm{tot}}}

\newcommand{\pT}{p_{\mathrm{T}}}

\newcommand{\pTj}{p_{\mathrm{T},\jet}}

\newcommand{\pTV}{p_{\mathrm{T},V}}

\newcommand{\beqar}{\begin{eqnarray}}
\newcommand{\eeqar}{\end{eqnarray}}
\newcommand{\beq}{\begin{equation}}
\newcommand{\eeq}{\end{equation}}
\newcommand{\bit}{\begin{itemize}}
\newcommand{\eit}{\end{itemize}}

\def\reffi#1{\mbox{Fig.~\ref{#1}}}

\newcommand{\NLO}{\mathrm{NLO}\xspace}
\newcommand{\QCD}{\mathrm{QCD}}

\newcommand{\QCDpEW}{\mathrm{QCD+EW}}
\newcommand{\QCDtEW}{\mathrm{QCD\times EW}}

\def\relplotwidth{0.29}

\newcommand{\Photo}{}

\usepackage{eso-pic}

\begin{document}

\AddToShipoutPictureBG*{%
  \AtPageUpperLeft{%
    \hspace{\paperwidth}%
    \raisebox{-3\baselineskip}{%
      \makebox[0pt][r]{DCPT/15/56, IPPP/15/28, MITP/15-037, ZU-TH 12/15  \hspace{2cm}}
}}}%

\vspace*{1cm}
\title{NLO QCD+EW automation and precise 
predictions for V+multijet production}

\author{S. Kallweit}

\address{Institut f\"ur Physik \& PRISMA Cluster of Excellence, 
Johannes Gutenberg Universit\"at, \\55099 Mainz, Germany}

\author{ J. M. Lindert\footnote{Speaker}, S. Pozzorini, M. Sch\"onherr }

\address{Physik-Institut, Universit\"at Z\"urich,
Winterthurerstrasse 190, \\
	CH-8057 Z\"urich,
	Switzerland}

\author{P.~Maierh\"ofer}

\address{Institute for Particle Physics Phenomenology, 
           Durham University, \\
           Durham DH1 3LE, 
           UK}

\maketitle\abstracts{
In this talk we present a fully automated implementation of next-to-leading order
electroweak (NLO EW) corrections in \OpenLoops together with \Sherpa and \Munich. 
As a first application, we present NLO QCD+EW predictions for 
the production of positively charged $W$ bosons in association with up to three jets 
and for the production of a $Z$ boson or photon in association with one jet.
}

\section{Introduction}

The upcoming Run-II of the LHC will probe the Standard Model (SM) of particle 
physics at unprecedented energies and precision.
At the TeV energy scale higher-order electroweak (EW) corrections can be strongly enhanced due to the 
presence of large Sudakov logarithms. Their inclusion in the experimental 
analyses will significantly enhance the sensitivity for new phenomena. 
Here we present a fully automated implementation of next-to-leading order (NLO) EW 
corrections, applicable to any process within the SM.

In the following, first we briefly review the recently accomplished automation of NLO
EW corrections in \OpenLoops~\cite{hepforge}, \Sherpa~\cite{Gleisberg:2008ta} and \Munich~\cite{munich}. 
Subsequently, we present numerical results for
W+multijet production and for Z-boson and photon production in conjunction
with one jet. Due to the large cross sections and clean experimental signatures these
processes represent an ideal laboratory to test the validity of theoretical methods and 
tools that are used for the simulation of a vast range of processes at the LHC.
Furthermore, they are important backgrounds for top- and Higgs-physics and various 
searches for physics beyond the Standard Model
including Dark Matter searches in the monojet channel.


%
Discussion and results presented here are partly based on
\cite{Kallweit:2014xda}, where more details can be found.

\section{NLO QCD+EW automation in \OpenLoops + \Sherpa/\Munich}

The calculation of NLO QCD corrections for any SM process was already well
established in the \OpenLoopsSherpaMunich programs.  This fully automated
framework has now been extended to NLO EW calculations.  More precisely, the
new implementation allows for NLO calculations at any given Order $\alpha_s^n
\alpha^m$, including all relevant QCD--EW interference effects.
Full NLO SM calculations that include all possible 
$\ord(\alpha_s^{n+k} \alpha^{m-k})$ contributions to a certain process
are also supported.

The \OpenLoops\cite{hepforge} program generates all relevant matrix-element ingredients, i.e.~one-loop amplitudes, tree amplitudes for Born and bremsstrahlung contributions, as well as colour-, charge-, gluon-helicity and photon-helicity correlations that are needed for infrared subtractions. 
The  \OpenLoops program is based on the Open Loops algorithm~\cite{Cascioli:2011va}, which employs a recursion to construct loops as tree structures supplemented with full loop-momentum information. 
Combined with the \Collier tensor reduction library \cite{Denner:2014gla} the employed recursion permits to achieve very high CPU performance and a high degree of numerical stability.

The kernel of the Open Loops recursion is universal and depends only on the Lagrangian of the model at hand. The  algorithm is thus applicable to any process within any renormalizable theory. The implementation has successfully been applied to various precision studies at NLO QCD and the extension to NLO electroweak corrections has very recently been achieved. It required the implementation of all $\ord(\alpha)$ EW Feynman rules  in the framework of the numerical Open Loops recursion  including counterterms associated with so-called $R_2$ rational parts~\cite{Garzelli:2009is} and with the on-shell renormalization of UV singularities~\cite{Denner:1991kt}. Additionally for the treatment of heavy unstable particles the complex mass scheme has been implemented. 
For the convenience of the user the \OpenLoops program is accompanied by a large process library including more than a hundred LHC processes -- currently all at the NLO QCD level but the library will be extented to NLO EW soon.

All complementary tasks, i.e.~the bookkeeping of partonic processes, the subtraction of IR singularities, and phase space integration, have been automated within \Munich and \Sherpa. Automated NLO EW simulations will be supported by future public releases of the employed tools.
%



\section{W+multijet production}
\label{sec:w_multijet}

\begin{figure*}[t]
\centering
   \includegraphics[width=\relplotwidth\textwidth]{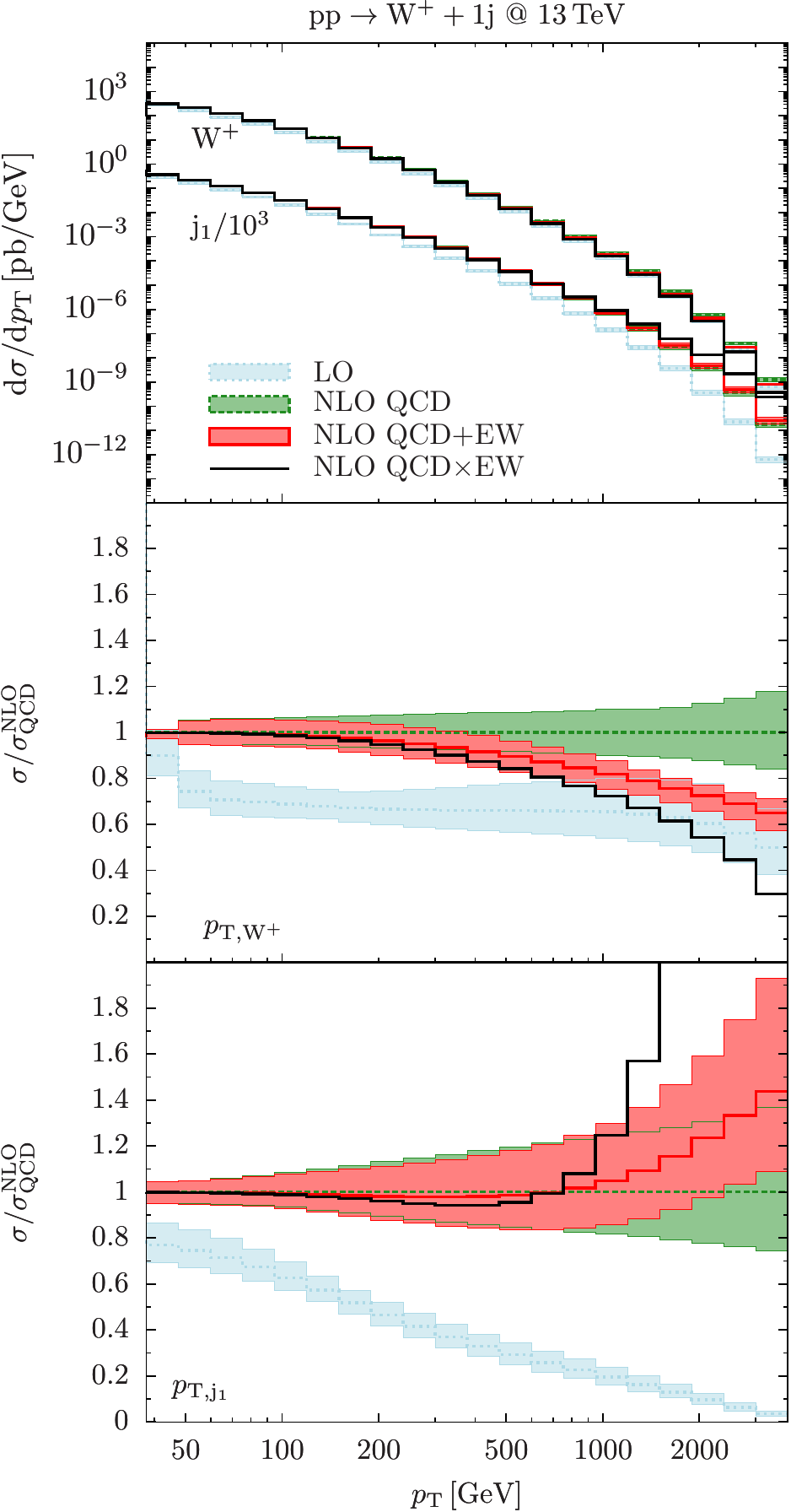}
   \includegraphics[width=\relplotwidth\textwidth]{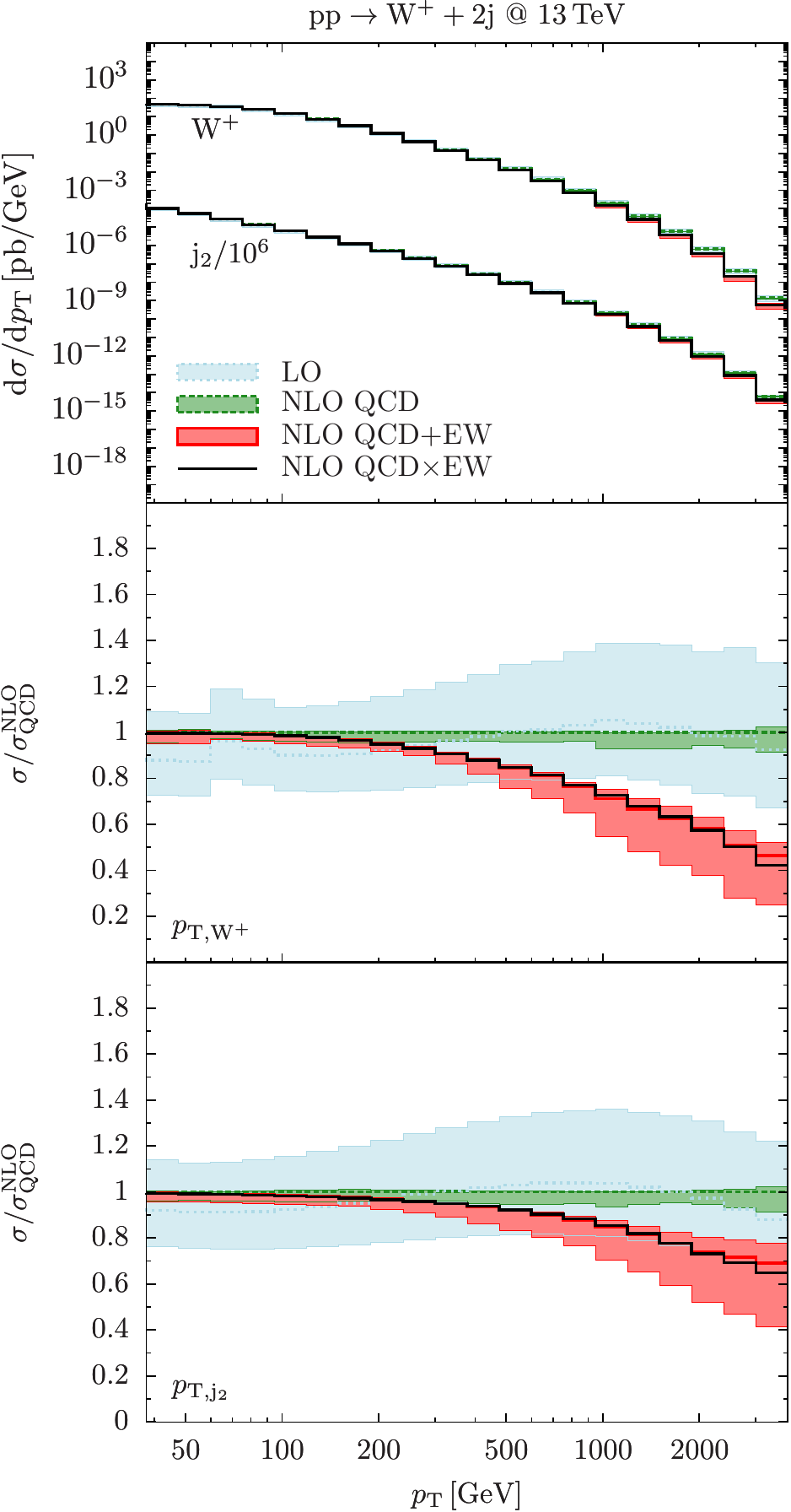}
   \includegraphics[width=\relplotwidth\textwidth]{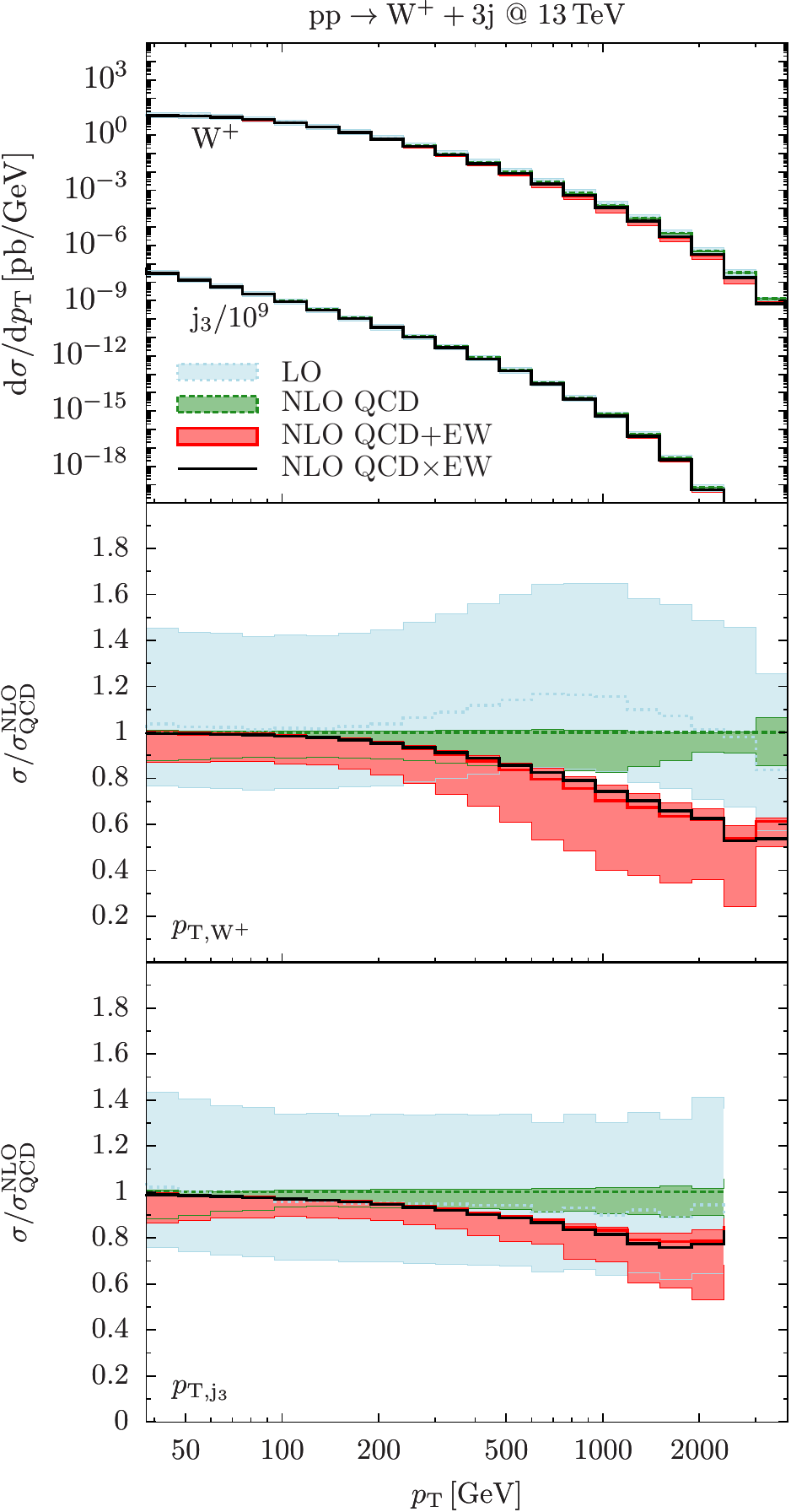}
\caption{
Distributions in the transverse momentum $\pT$ of the $W^+$ and the n-th jet for inclusive \mbox{$W^++n\jet$} production with \mbox{$n=1$ (left), $n=2$ (center), $n=3$ (right)} at $\sqrt{s}=13$~TeV.
Absolute LO (light blue), NLO QCD (green), NLO QCD+EW (red) and NLO
QCD$\times$EW (black) predictions (upper panel) and relative corrections
with respect to NLO QCD (lower panels).  
The bands correspond to scale variations, 
and in the case of relative corrections only the numerator is varied. 
}
\label{fig:wjets_pTall}
\end{figure*}

As a first highly non-trivial application we study the production of $W^+ + n$
jets with $n=1,2,3$ at the LHC including NLO QCD and NLO EW corrections. 
In Fig.~\ref{fig:wjets_pTall} we show differential distribution in the transverse
momentum of the produced $W^+$ and the n-th jet. We use the anti-$k_T$ jet clustering 
algorithmn with R=0.4 and require 
$\pT>30$ GeV and $|\eta| < 4.5$ for the jets.
Besides LO predictions of $\ord(\alpha_S^n \alpha)$ we show $\NLO~\QCD$ predictions including corrections of
$\ord(\alpha_S^{n+1} \alpha)$ and 
NLO $\QCDpEW$ predictions including $\ord(\alpha_S^{n+1}\alpha)+\ord(\alpha_S^n \alpha^2)$ corrections.
Theoretical uncertainties are assessed by standard variations
of the renormalization and factorization scales.\cite{Kallweit:2014xda}
Besides $\QCDpEW$ predictions we also
show factorized $\QCDtEW$ predictions were the $\NLO~\QCD$ predictions are multiplied with 
a NLO EW K-factor.  A difference between the two approaches indicates uncertainties 
due to missing two-loop EW-QCD corrections. In the lower panel we show corrections with respect to the NLO QCD prediction.
In the tail of the $\pT$ distribution of the jet in $W+1$ jet production the NLO QCD corrections grow larger then a factor of 10 and the EW corrections turn positive. Together with the large scale uncertainties this is a clear indication for a poor perturbative convergence. Indeed, NLO corrections to inclusive $W+1$ jet production are dominated by dijet configurations radiating a relatively soft $W$,
which are effectively of leading order. Such configurations appear 
already at LO for $W+2$ jet production (shown in the central plot), where the NLO QCD corrections are small and stable. Here, the EW corrections show a typical Sudakov behaviour and reach $-30(-60)\%$ and $-15(-25)\%$ at $1(4)$ TeV for the $\pT$ of the $W^+$ and the 2nd jet respectively. A similar picture emerges for  $W+3$ jet production (shown in the right plot). 

\begin{figure*}[t]
\centering
   \includegraphics[width=\relplotwidth\textwidth]{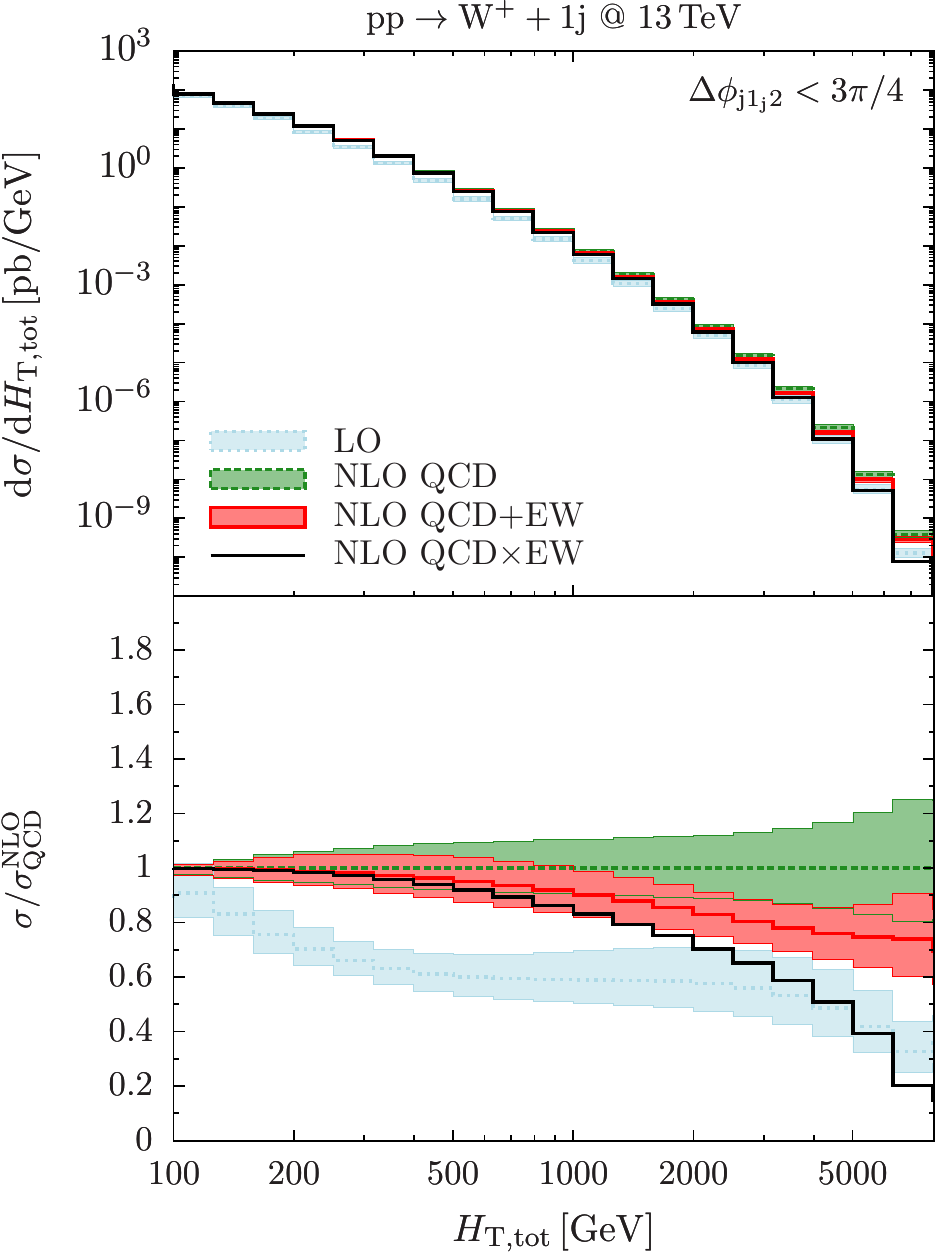}
   \includegraphics[width=\relplotwidth\textwidth]{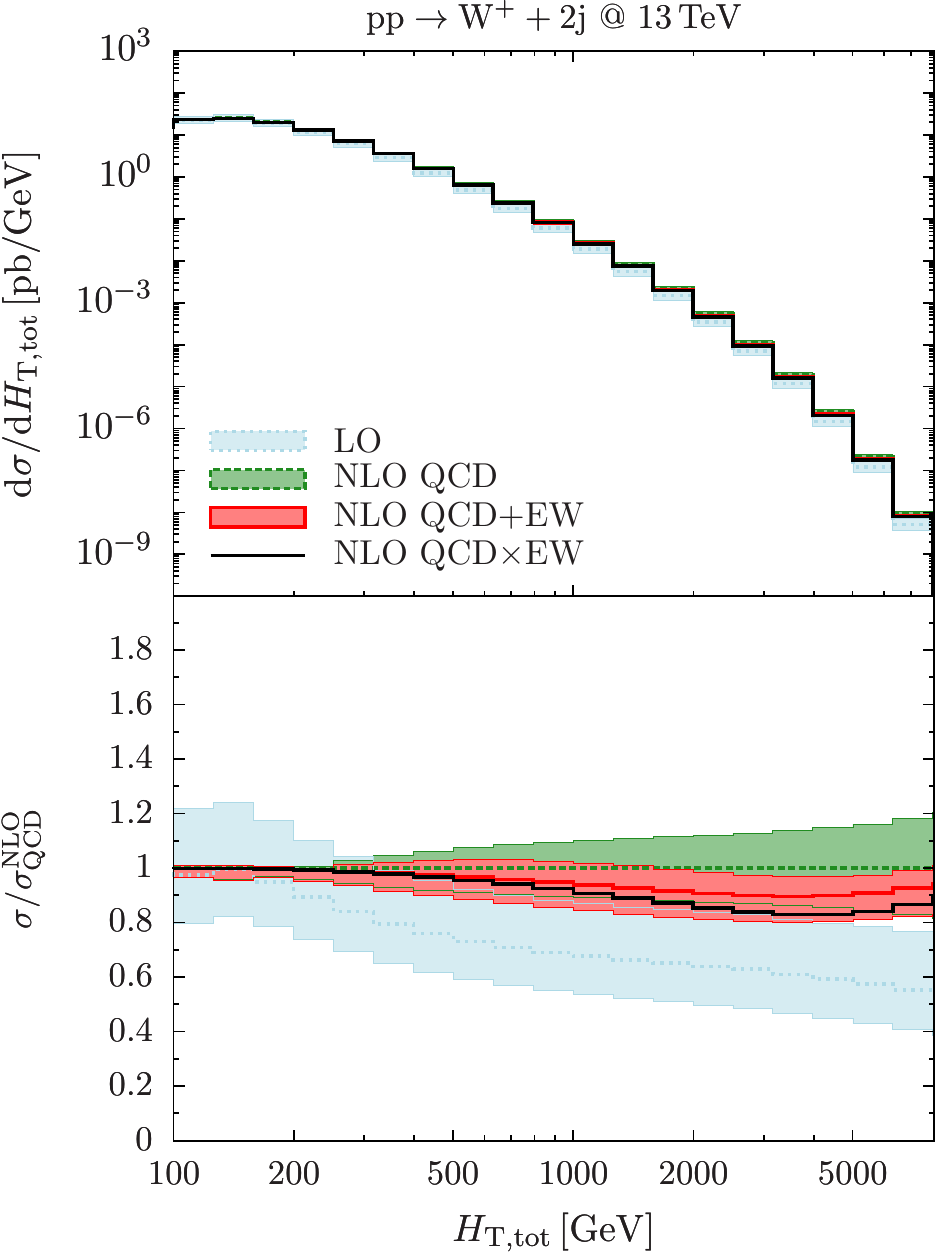}
   \includegraphics[width=\relplotwidth\textwidth]{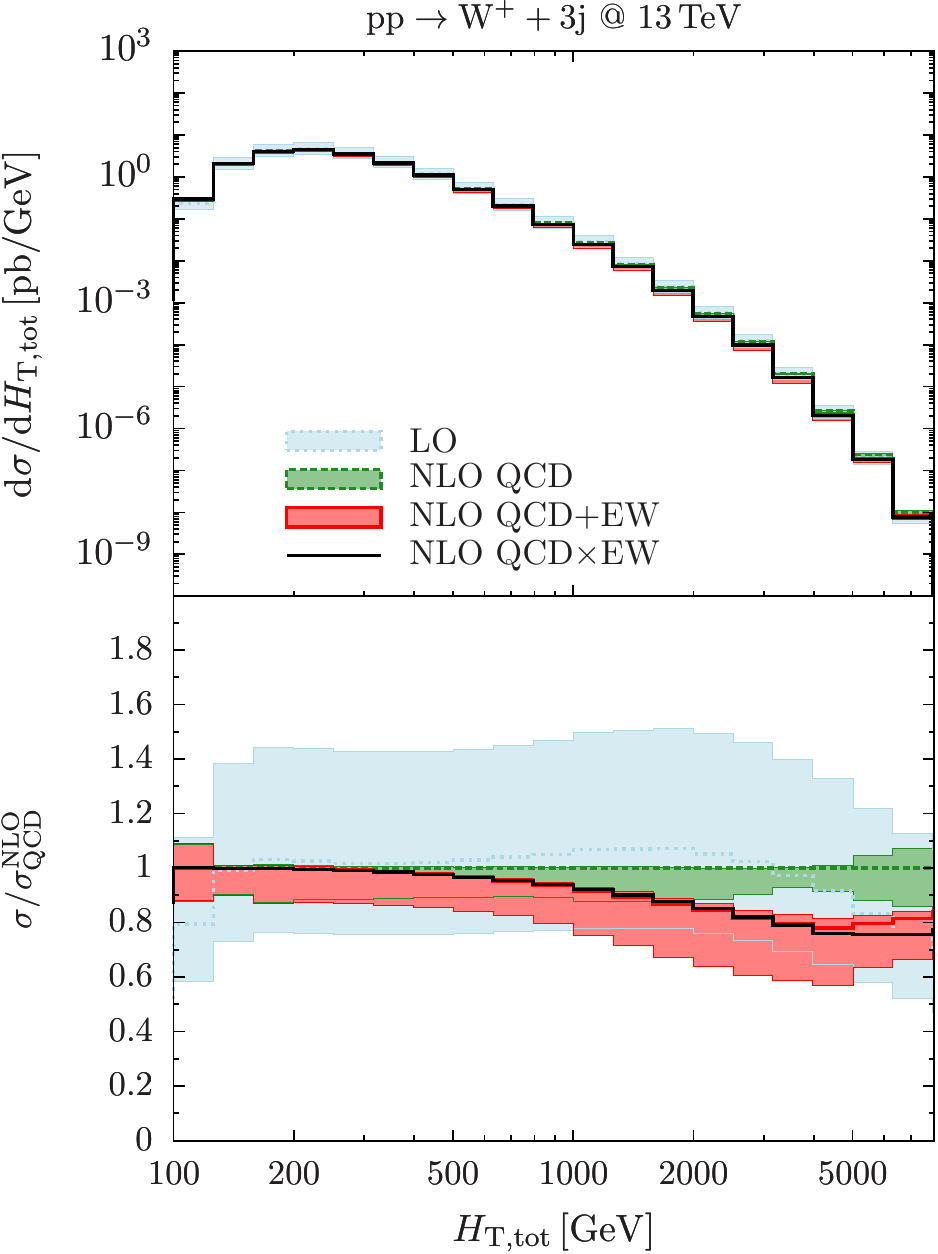}
\caption{
Distributions in $\HTtot$ for exclusive \mbox{$W^++1\jet$} (left), inclusive \mbox{$W^++2\jet$} (center)
and  inclusive \mbox{$W^++3\jet$} (right) production.
Curves and bands as in \reffi{fig:wjets_pTall}.
}
\label{fig:wjets_HTtot}
\end{figure*}

In Fig. \ref{fig:wjets_HTtot} we show differential distributions in $\HTtot$ -- the scalar sum of all final state transverse 
momenta for $W^+ + 1,2,3$j production. In order to improve the perturbative convergence in the case of $W^+ + 1$j production we employ a veto on a second jet if $\Delta\phi_{j_1j_2} > 3/4\pi$. Still, for very large $\HTtot$ the QCD corrections to $W^+ + 1$j and $W^+ + 2$j  production increase strongly, suppressing the impact of the EW corrections. Only for $W^+ + 3$j production the QCD corrections are stable in all of the considered range. However, here the EW corrections are still moderate and only increase beyond the NLO QCD scale uncertainties for $\HTtot$  at the TeV scale.

\section{ Z/$\gamma$+jet production}
\label{sec:z_gamma_jet}

As a second application we study the production of a Z boson or a photon in association with a jet at the LHC with $\sqrt{s}=8$~TeV. The ratio of these processes differential in the $\pT$ of the produced gauge bosons can be used to model $Z(\to\nu\bar\nu)+1j$ production from a precise experimental measurement of $\gamma+1j$ production, i.e.~the dominant irreducible SM background in monojet searches for Dark Matter.

\begin{figure*}[t]
\centering
   \includegraphics[width=\relplotwidth\textwidth]{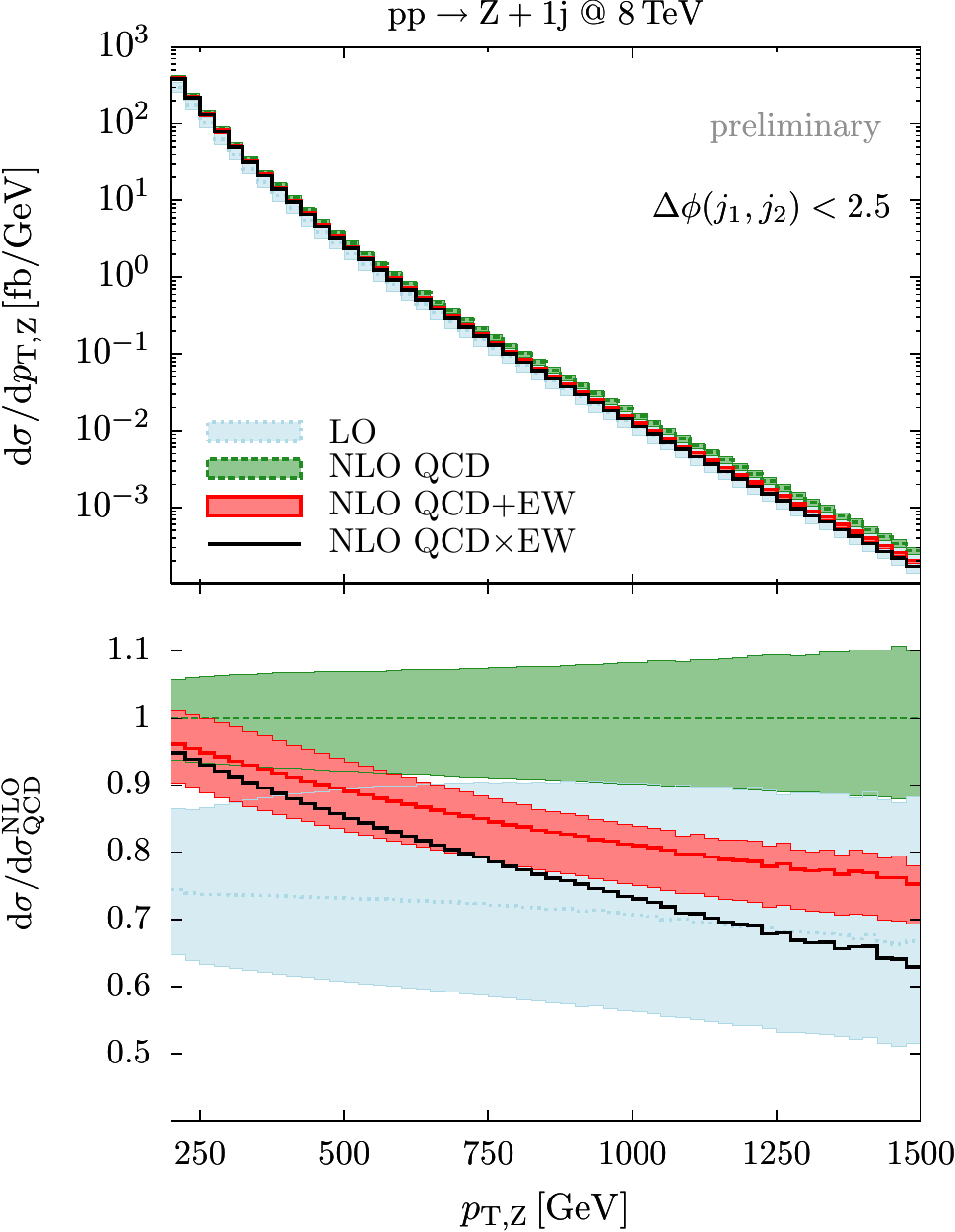}
   \includegraphics[width=\relplotwidth\textwidth]{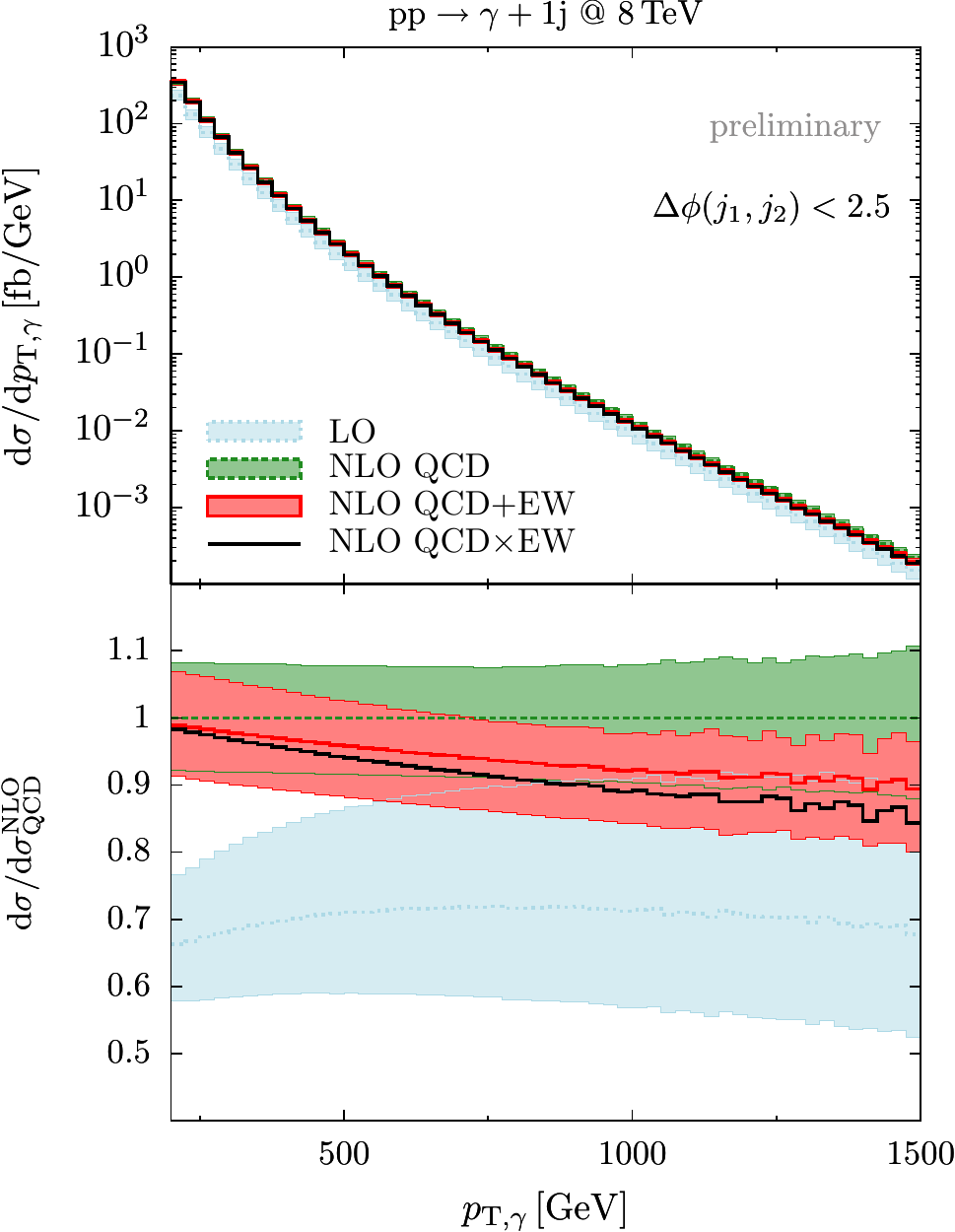}
       \includegraphics[width=\relplotwidth\textwidth]{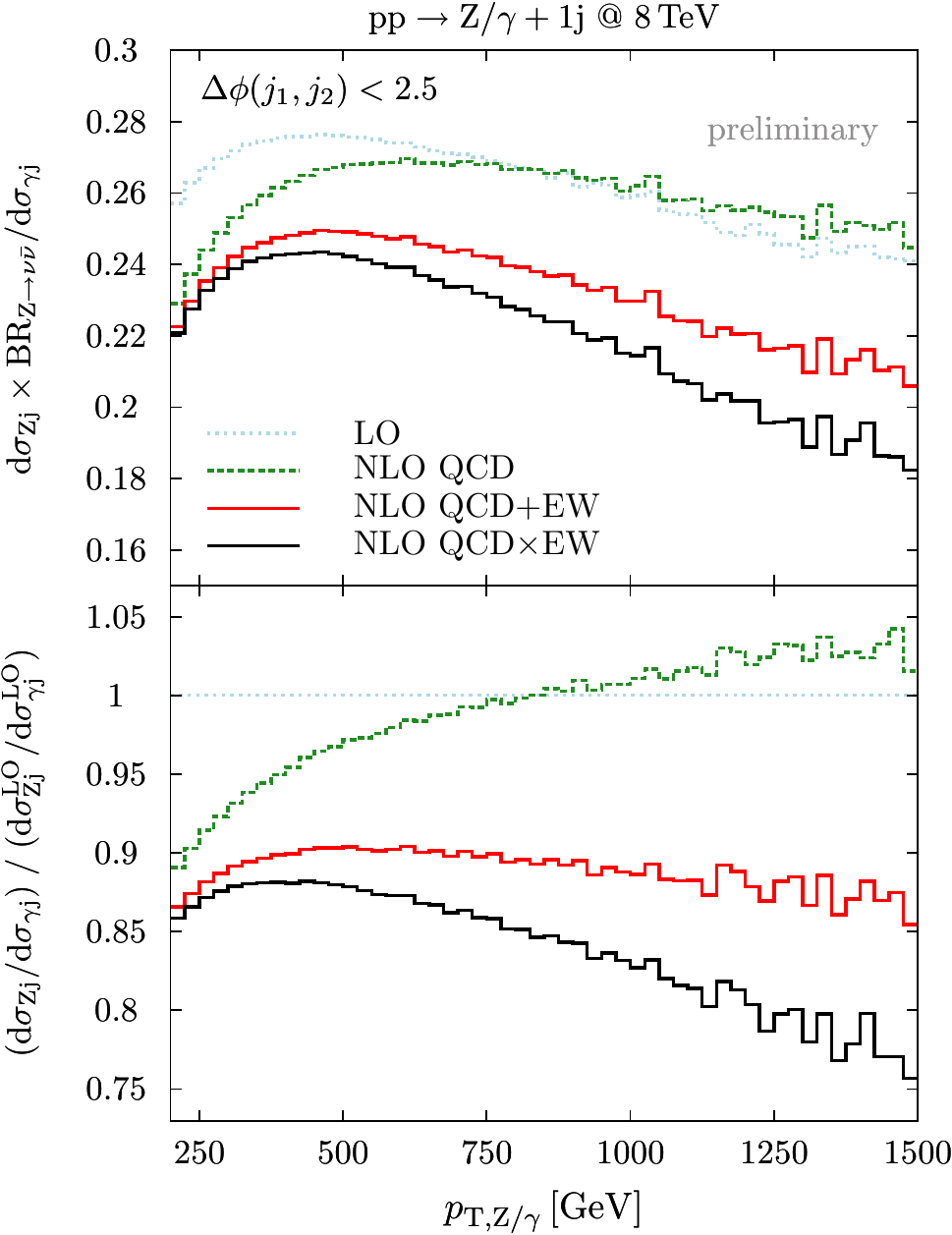}
\caption{
Distributions in the transverse momenta of the $Z$ boson (left) and of the photon (center)
for \mbox{$Z+1\jet$} and \mbox{$\gamma+1\jet$} production at $\sqrt{s}=8$~TeV.
Curves and bands as in \reffi{fig:wjets_pTall}.
In the right plot the ratio of the  $\pT$ of the Z and the photon together with the 
 relative corrections in the ratio with respect to the LO ratio are shown using
the same color coding as before. 
}
\label{fig:zgammaj_pT}
\end{figure*}

Results in the $\pT$ of the produced weak gauge boson are shown in Fig.~\ref{fig:zgammaj_pT} with the same color coding and nomenclature as before. In the left and central plot results for $Z+1j$  and $\gamma+1j$ production are shown respectively. We require for the associated jet $\pTj > 110$~GeV and $|\eta_j| <2.4$ and veto a possible second jet with $\pTj > 30$~GeV and $\Delta\phi_{j_1j_2} > 2.5$.
These cuts are in agreement with a setup employed by CMS in an upcoming monojet search.
The NLO QCD corrections to both processes are almost identical at large transverse momentum of the produced gauge bosons $\pTV$, while they differ slightly at small $\pTV$ due to the finite mass of the produced $Z$. NLO QCD scale uncertainties are at the level of $10\%$. On the contrary, the EW corrections to $Z+1j$ production are enhanced compared to $\gamma+1j$ production and at 1~TeV they reach $-20\%$ and $-8\%$ respectively. In the right plot of Fig.~\ref{fig:zgammaj_pT} we show the ratio in $\pTV$ of  $Z+1j$ over  $\gamma+1j$ production. 
This observable is fairly stable in the considered $\pT$ range and QCD corrections are below $10\%$.
However, EW corrections result in an almost constant shift of about $10\%$ comparing the $\pT$-ratio at LO and NLO $\QCDpEW$. Such a shift is consistent with the observed deviation presented by CMS at Moriond 2015 QCD (also shown in Fig.~6 of~\cite{CMS:2014fha}).

\section{Conclusions}
Recent progress in the automation of perturbative calculations
within the \OpenLoops+\Munich/\Sherpa frameworks has opened the door to NLO QCD+EW simulations for a
vast range of Standard Model processes, up to high particle multiplicity, at
current and future colliders.
The large impact of NLO EW effects in $V+$multijet production at high energy 
demonstrates the relevance of these new tools for the upcoming Run-II of the LHC.

%


%



\section*{References}

\begin{thebibliography}{99}



\bibitem{hepforge}
The {\sc OpenLoops} one-loop generator by F.~Cascioli, J.~Lindert,
  P.~Maierh{\"o}fer and S.~Pozzorini is publicly available at {\tt
  http://openloops.hepforge.org}.


\bibitem{Gleisberg:2008ta}
T.~Gleisberg {\em et~al.}, 
  \href{http://dx.doi.org/10.1088/1126-6708/2009/02/007}{{\em JHEP} {\bfseries
  0902} (2009) 007},
[\href{http://arxiv.org/abs/0811.4622}{{\ttfamily arXiv:0811.4622}}].


\bibitem{munich}
{\sc Munich}---an automated parton level NLO generator by S.~Kallweit. In
  preparation.


\bibitem{Kallweit:2014xda}
S.~Kallweit {\em et~al.},
[\href{http://arxiv.org/abs/1412.5157}{{\ttfamily arXiv:1412.5157}}].

\bibitem{Cascioli:2011va}
F.~Cascioli {\em et~al.}, \href{http://dx.doi.org/10.1103/PhysRevLett.108.111601}{{\em
  Phys.Rev.Lett.} {\bfseries 108} (2012) 111601},
[\href{http://arxiv.org/abs/1111.5206}{{\ttfamily arXiv:1111.5206}}].





\bibitem{Denner:2014gla}
A.~Denner, S.~Dittmaier, and L.~Hofer, [\href{http://xxx.lanl.gov/abs/1407.0087}{{\tt
 arXiv:1407.0087}}].

\bibitem{Garzelli:2009is}
M.~Garzelli {\em et~al.},  {\em JHEP} {\bf 1001} (2010)
  040, [\href{http://xxx.lanl.gov/abs/0910.3130}{{\tt arXiv:0910.3130}}].

\bibitem{Denner:1991kt}
A.~Denner,   {\em Fortsch.Phys.} {\bf 41} (1993) 307--420,
  [\href{http://xxx.lanl.gov/abs/0709.1075}{{\tt arXiv:0709.1075}}].

\bibitem{CMS:2014fha} 
  CMS Collaboration [CMS Collaboration],
  CMS-PAS-SMP-14-005.

\end{thebibliography}
\end{document}